\documentclass{iau307}
\usepackage{graphicx}
\usepackage{natbib}
\usepackage{url}
\usepackage{dtklogos}
\usepackage{gensymb}
\bibpunct{(}{)}{;}{a}{}{,}

\title[Beam me up, Spotty] 
{Beam me up, Spotty: Toward a new understanding of the physics of massive star photospheres}

\author[David-Uraz, Wade \& Owocki]   
{Alexandre David-Uraz$^1$, Gregg Wade$^2$ \and Stan Owocki$^3$}

\affiliation{$^1$Queen's University, Canada \\ email: {\tt adavid-uraz@astro.queensu.ca} \\[\affilskip]
$^2$Royal Military College, Canada \\[\affilskip]
$^3$University of Delaware, USA}

\pubyear{2014}
\volume{307} 
\pagerange{}
\jname{New windows on massive stars: asteroseismology, interferometry, and spectropolarimetry}
\editors{G. Meynet, C. Georgy, J.H. Groh \& Ph. Stee, eds.}

\begin{document}

\maketitle

\begin{abstract}
For 30 years, cyclical wind variability in OB stars has puzzled the astronomical community. Phenomenological models
involving co-rotating bright spots provide a potential explanation for the observed variations, but the underlying physics remains unknown. We present recent results
from hydrodynamical simulations constraining bright spot properties and compare them to what can be inferred from 
space-based photometry. We also explore the possibility that these spots are caused by magnetic fields and discuss the detectability of such fields.
\keywords{Hydrodynamics, methods: numerical, stars: winds, outflows}
\end{abstract}

\firstsection 
\section{Introduction}

Massive stars exhibit cyclical spectral variability on timescales of hours to days. One seemingly ubiquitous manifestation of these variations
is the presence of so-called ``discrete absorption components" (DACs, \citealt{1989ApJS...69..527H}): additional absorption features
within the P Cygni absorption troughs of wind-sensitive UV resonance lines which progress from low to near-terminal velocity. They are
believed to be rotationally modulated.
The generally-accepted hypothesis which explains their formation is the ``co-rotating interaction regions" model (CIR, \citealt{1986A&A...165..157M}).
Bright spots on the surface of the star locally drive an enhanced wind, leading to spiral-like structures (\citealt{1996ApJ...462..469C}, henceforth
referred to as CO96). 
The most important feature of this model is the presence of ``velocity kinks" in the wind; the driving of extra material by the bright spots
cannot be sustained (the wind is overloaded) and that material 
eventually decelerates, which results in a velocity plateau, thus leading to an accumulation of matter at a given
velocity which causes the extra absorption associated with DACs.

\section{Modern constraints and revisited model\label{SecOne}}

CO96 provide a phenomenological explanation for the DAC phenomenon. 
Consequently, the presence of bright spots was not \textit{a priori} physically motivated (nor constrained). However,
with the advent of space-based photometry (e.g. the MOST space telescope), the precisely-determined amplitude of light-curve variations places very strong
constraints on the size and contrast of putative spots on the surface of the star.

\begin{figure}[b]
\begin{center}
\includegraphics[width=\textwidth]{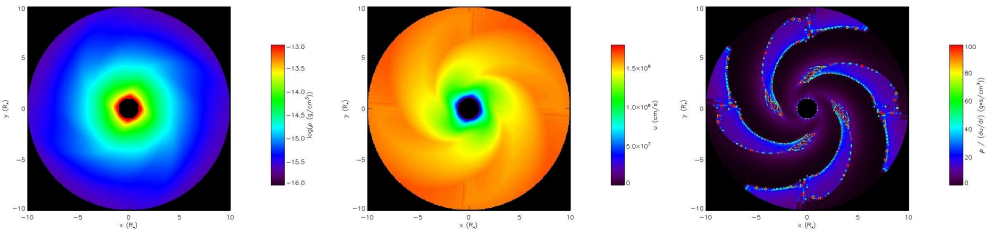} 
\caption{Preliminary 2D hydrodynamic models show that four spots with $10\degree$ radius and 33\% brightness enhancement 
allow us to qualitatively reproduce the conditions necessary to obtain
DACs. From left to right: density profile, radial velocity profile, density divided by the radial velocity gradient (this proxy variable acts like the
optical depth and the bright trailing edges indicate the presence of velocity kinks).}
\label{fig1}
\end{center}
\end{figure}

\citet{2014MNRAS.441..910R} find a 10 mmag amplitude for $\xi$ Per, constraining any spots present to be much more modest than those used by CO96. 
It is unclear whether velocity kinks can be formed with such modest spots. 
One possible way to facilitate kink formation is by including ionization effects, which were not taken into
account by CO96. Following up on that work, we aim to study the effects of adding the ionization correction ($\delta$) in hydrodynamic wind simulations, 
using the VH-1 package.
1D simulations show that for high values of the $\delta$ parameter \citep{1982ApJ...259..282A}, 
the wind becomes overloaded and we 
obtain a coasting solution, even without spots. While this work is quite preliminary, 2D hydrodynamical simulations based on realistic stellar parameters
(in this case, $\xi$ Per, with $10\degree$ spots and a 33\% brightness enhancement) and including additional wind physics (ionization effect, finite disk
correction) show that we can reproduce CIRs and even obtain the desired velocity kinks (see Fig.~\ref{fig1}).

\section{The role of magnetic fields\label{SecBfield}}

Magnetic fields constitute a popular hypothesis to explain the surface nonuniformities apparently needed to form CIRs. 
While only a small fraction of OB stars are known to host detectable fields (about 7\%, \citealt{2013arXiv1310.3965W}), most known magnetic
massive stars have a large scale, essentially dipolar field. However, \citet{2014arXiv1407.6417D} 
effectively rule out large-scale magnetic fields as the physical cause of DACs.
Recent models (e.g. \citealt{2009A&A...499..279C}) suggest the existence of a sub-surface convection layer in massive stars 
which could produce small-scale surface magnetic fields. \citet{2013A&A...554A..93K} have characterized the detectability 
of randomly distributed magnetic spots. For $v \sin i < 50$ km/s, one would expect to detect $10\degree$ magnetic spots
with a field strength of over 100 G (with a 0.5 filling factor); 
to produce a 10 mmag photometric variation, such a spot should have a field strength of about 330 G
(using the formula of \citealt{2014arXiv1407.6417D}), which should therefore
be detectable with current instrumentation. 
Further work on both the numerical and observational fronts will be required to better understand and constrain the role of small-scale 
magnetic fields in producing DACs.


\bibliographystyle{iau307}
\bibliography{MyBiblio}

\end{document}